\documentclass[11pt]{article}  
\usepackage{amsmath}
\usepackage{amsfonts,amssymb}

\begin{document}

\newtheorem{theorem}{Theorem}[section]
\newtheorem{lemma}{Lemma}[section]
\newcommand{\bproof}{\smallskip{\em Proof. }}
\newcommand{\eproof}{\smallskip}
\newcommand{\qed}{{\hbox{$\  \square$}}}

\newcommand{\nl}{\nonumber\\}
\newcommand{\nnl}{\nl[6mm]}
\newcommand{\nle}{\nl[-2.0mm]\\[-2.0mm]}
\newcommand{\nlb}[1]{\nl[-2.0mm]\label{#1}\\[-2.0mm]}
\newcommand{\bl}{&&\quad}
\newcommand{\ab}{\allowbreak}

\renewcommand{\leq}{\leqslant}
\renewcommand{\geq}{\geqslant}

\renewcommand{\theequation}{\thesection.\arabic{equation}}
\let\ssection=\section
\renewcommand{\section}{\setcounter{equation}{0}\ssection}

\newcommand{\be}{\bes}
\newcommand{\ee}{\ees}
\newcommand{\bes}{\begin{eqnarray}}
\newcommand{\ees}{\end{eqnarray}}
\newcommand{\eens}{\nonumber\end{eqnarray}}
\newcommand{\barr}{\begin{array}}
\newcommand{\earr}{\end{array}}

\renewcommand{\/}{\over}
\renewcommand{\d}{\partial}
\newcommand{\ddt}{{d\/dt}}
\newcommand{\dNx}{d^N\!x}
\newcommand{\dxdt}{\dNx\,dt}
\newcommand{\mx}{m\kern-0.5mm\cdot\kern-0.5mm x}
\newcommand{\mq}{m\kern-0.5mm\cdot\kern-0.5mm q}

\newcommand{\eps}{\epsilon}
\newcommand{\dlt}{\delta}
\newcommand{\al}{\alpha}
\newcommand{\bt}{\beta}
\newcommand{\gm}{\gamma}
\newcommand{\z}{\zeta}
\newcommand{\si}{\sigma}
\newcommand{\la}{\lambda}
\newcommand{\La}{\Lambda}
\newcommand{\ka}{\kappa}
\newcommand{\emp}{\emptyset}

\newcommand{\xmu}{\xi^\mu}
\newcommand{\xnu}{\xi^\nu}
\newcommand{\xz}{\xi^0}
\newcommand{\ynu}{\eta^\nu}
\newcommand{\yz}{\eta^0}
\newcommand{\dmu}{\d_\mu}
\newcommand{\dnu}{\d_\nu}
\newcommand{\dsi}{\d_\si}
\newcommand{\dtau}{\d_\tau}
\newcommand{\drho}{\d_\rho}
\newcommand{\qmu}{q^\mu}
\newcommand{\qnu}{q^\nu}
\newcommand{\qsi}{q^\si}
\newcommand{\qtau}{q^\tau}
\newcommand{\qrho}{q^\rho}
\newcommand{\pmu}{p_\mu}
\newcommand{\pnu}{p_\nu}
\newcommand{\prho}{p_\rho}

\newcommand{\dsum}[2]{\sum_{\scriptstyle{#1}\atop\scriptstyle{#2}}}

\newcommand{\summ}[1]{\sum_{|\mm|\leq #1}}
\newcommand{\sumnm}{\sum_{|\nn|\leq |\mm|}}
\newcommand{\summn}[1]{\sum_{|\nn|\leq |\mm|\leq #1}}

\newcommand{\xxy}{\xi\leftrightarrow\eta}
\newcommand{\fxg}{f\leftrightarrow g}

\newcommand{\fa}{\phi_\alpha}
\newcommand{\fb}{\phi_\beta}
\newcommand{\pa}{\pi^\alpha}
\newcommand{\pb}{\pi^\beta}

\newcommand{\Ea}{\EE^\alpha}
\newcommand{\Eb}{\EE^\beta}

\newcommand{\Dslash}{\hbox{$D\kern-2.4mm/\,$}}

\newcommand{\fs}{\phi^*}
\newcommand{\fsa}{\phi^{*\alpha}}
\newcommand{\fsb}{\phi^{*\beta}}
\newcommand{\psa}{\pi^*_\alpha}
\newcommand{\psb}{\pi^*_\beta}

\renewcommand{\fam}{\phi_{\alpha,\mm}}
\newcommand{\fan}{\phi_{\alpha,\nn}}
\newcommand{\pam}{\pi^{\alpha,\mm}}
\newcommand{\pan}{\pi^{\alpha,\nn}}

\newcommand{\fbm}{\phi_{\beta,\mm}}
\newcommand{\fbn}{\phi_{\beta,\nn}}
\newcommand{\pbm}{\pi^{\beta,\mm}}
\newcommand{\pbn}{\pi^{\beta,\nn}}

\newcommand{\wfa}{{\overline\phi}_\alpha}
\newcommand{\wpa}{{\overline\pi}{}^\alpha}
\newcommand{\wfsa}{{\overline\phi}{}^{*\alpha}}
\newcommand{\wpsa}{{\overline\pi}{}^*_\alpha}
\newcommand{\wfam}{{\overline\phi}_{\alpha,\mm}}
\newcommand{\wfbn}{{\overline\phi}_{\beta,\nn}}
\newcommand{\wpam}{{\overline\pi}{}^{\alpha,\mm}}
\newcommand{\wfsam}{{\overline\phi}{}^{*\alpha}_{,\mm}}
\newcommand{\wfsan}{{\overline\phi}{}^{*\alpha}_{,\nn}}
\newcommand{\wfsbn}{{\overline\phi}{}^{*\beta}_{,\nn}}
\newcommand{\wpsam}{{\overline\pi}{}^{*,\mm}_\alpha}
\newcommand{\wpsan}{{\overline\pi}{}^{*,\nn}_\alpha}
\newcommand{\wpsbn}{{\overline\pi}{}^{*,\nn}_\beta}
\newcommand{\wpsbm}{{\overline\pi}{}^{*,\mm}_\beta}
\newcommand{\wba}{{\overline b}{}^a}
\newcommand{\wca}{{\overline c}_a}
\newcommand{\wbam}{{\overline b}{}^a_{,\mm}}
\newcommand{\wban}{{\overline b}{}^a_{,\nn}}
\newcommand{\wbbn}{{\overline b}{}^b_{,\nn}}
\newcommand{\wcam}{{\overline c}{}^{,\mm}_a}
\newcommand{\wcbm}{{\overline c}{}^{,\mm}_b}

\newcommand{\fsam}{\phi^{*\alpha}_{,\mm}}
\newcommand{\fsan}{\phi^{*\alpha}_{,\nn}}
\newcommand{\fsbn}{\phi^{*\beta}_{,\nn}}
\newcommand{\psam}{\pi^{*,\mm}_\alpha}
\newcommand{\psbm}{\pi^{*,\mm}_\beta}
\newcommand{\psbn}{\pi^{*,\nn}_\beta}
\newcommand{\Eam}{\EE^\alpha_{,\mm}}
\newcommand{\bam}{b^a_{,\mm}}
\newcommand{\ban}{b^a_{,\nn}}
\newcommand{\bbn}{b^b_{,\nn}}
\newcommand{\cam}{c^{,\mm}_a}
\newcommand{\cbm}{c^{,\mm}_b}
\newcommand{\ram}{r^a_{,\mm}}

\newcommand{\half}{{1\/2}}
\newcommand{\Np}[1]{{N+p\choose N #1}}
\newcommand{\Npr}{{N+p-r\choose N-r}}

\newcommand{\mm}{{\mathbf m}}
\newcommand{\nn}{{\mathbf n}}
\newcommand{\rr}{{\mathbf r}}
\renewcommand{\ss}{{\mathbf s}}
\newcommand{\uu}{{\mathbf u}}
\newcommand{\vv}{{\mathbf v}}
\newcommand{\xx}{{\bar x}}
\newcommand{\one}{{\hat 1}}
\newcommand{\two}{{\hat 2}}

\renewcommand{\L}{{\cal L}}
\newcommand{\J}{{\cal J}}
\newcommand{\II}{{\cal I}}
\newcommand{\GG}{{\cal G}}
\newcommand{\UU}{{\cal U}}
\newcommand{\QQ}{{\cal Q}}
\newcommand{\PP}{{\cal P}}
\newcommand{\EE}{{\cal E}}
\newcommand{\FF}{{\cal F}}
\newcommand{\Lxi}{\L_\xi}
\newcommand{\Leta}{\L_\eta}

\newcommand{\into}{\hookrightarrow}
\newcommand{\e}{{\rm e}}
\newcommand{\ext}{{\rm ext\,}}
\newcommand{\afn}{{\rm afn\,}}
\newcommand{\larroww}[1]{{\ \stackrel{#1}{\longleftarrow}\ }}
\renewcommand{\th}{^{{\rm th}}}

\newcommand{\ord}{o}
\newcommand{\ordg}{\varsigma}

\newcommand{\repi}{\rep^{(i)}}
\newcommand{\Mi}{M^{(i)}}
\newcommand{\mri}{(-)^i {r\choose i}}

\newcommand{\dzero}{d^{(0)}}
\newcommand{\done}{d^{(1)}}
\newcommand{\di}{d^{(i)}}

\newcommand{\vect}{{\mathfrak{vect}}}
\newcommand{\map}{{\mathfrak{map}}}
\newcommand{\mb}{{\mathfrak{mb}}}
\newcommand{\vle}{{\mathfrak{vle}}}

\newcommand{\rep}{{\varrho}}

\newcommand{\tr}{{\rm tr}\kern0.7mm}
\newcommand{\oj}{{\mathfrak g}}
                                            
\renewcommand{\gg}{\oj\oplus gl(N)}
\newcommand{\dmap}{\vect(N)\ltimes \map(N,\oj)}

\newcommand{\bra}[1]{\big{\langle}#1\big{|}}
\newcommand{\ket}[1]{\big{|}#1\big{\rangle}}

\newcommand{\no}[1]{{\,:\kern-0.7mm #1\kern-1.2mm:\,}}

\newcommand{\RR}{{\mathbb R}}
\newcommand{\CC}{{\mathbb C}}
\newcommand{\ZZ}{{\mathbb Z}}
\newcommand{\NN}{{\mathbb N}}

\title{{Koszul-Tate Cohomology as Lowest-Energy Modules of Non-Centrally
Extended Diffeomorphism Algebras }}

\author{T. A. Larsson \\
Vanadisv\"agen 29, S-113 23 Stockholm, Sweden\\
email: thomas.larsson@hdd.se}

\maketitle
\begin{abstract}
Fock modules for multi-dimensional Virasoro algebras (non-central 
extensions of the diffeomorphism algebra $\vect(N)$) have recently 
been reported. Using ideas from the antifield formalism, I construct new 
classes of lowest-energy modules, as cohomology groups of a certain
Fock complex.
The Fock construction involves a passage to $p$-jets prior to normal
ordering, but the abelian charges usually diverge in the limit 
$p\to\infty$. The requirement of a finite limit imposes severe 
restrictions on the number of spacetime dimensions and on the order of 
the Euler-Lagrange (EL) equations. Under some natural assuptions 
(the EL equations are first order for fermions and second order for
bosons, and no reducible gauge symmetries appear), finiteness is only 
possible when the number of spacetime dimensions $N=4$.
\end{abstract}

\renewcommand{\arraystretch}{1.4}

\section{Introduction}
In a recent paper \cite{Lar98}, I constructed Fock modules of 
non-centrally extended diffeomorphism and current algebras in 
$N$-dimensional spacetime, i.e. the higher-dimensional generalizations
of Virasoro and affine algebras. More precisely, I considered the
{\em DGRO (Diffeomorphism, Gauge, Repara\-metri\-zation, Observer)} 
algebra $DGRO(N,\oj)$, where $\oj$ is a finite-dimensional Lie algebra. 
The crucial idea was to first expand all 
fields in a multi-dimensional Taylor series around the points on a 
one-dimensional curve (``the observer's trajectory''), and then to 
truncate at some finite order $p$. We thus obtain a realization of 
$\vect(N)\ltimes \map(N,\oj)$ (semi-direct product of diffeomorphism and
current algebras)\footnote{In previous writings, I have denoted 
the diffeomorphism algebra, or algebra of vector fields,
$\vect(N)$ by $diff(N)$.} on the space of trajectories in the space 
of tensor-valued
$p$-jets. This space consists of finitely many functions of a single
variable, which is precisely the situation where the normal ordering
prescription works. After normal ordering, a Fock
representation of the DGRO algebra is obtained.
Related work can be found in 
\cite{BB98,BBS01,Bil97,Lar97,Lar98,Lar99,MRY90,RMY92,RM94}.
Cocycles of the diffeomorphism algebra were classified by
Dzumadil'daev \cite{Dzhu96} and reviewed in \cite{Lar00}.

To progress further, one now wants to construct more interesting modules
of lowest-energy type. A natural idea is to consider a complex of Fock
modules:
\be
\barr{ccccccccccc}
... & \larroww Q & J^p\FF^{-1} & \larroww Q & J^p\FF^{0} & \larroww Q 
& J^p\FF^{1} & \larroww Q & J^p\FF^{2} & \larroww Q & ... \\
&& \downarrow \L && \downarrow \L 
&& \downarrow \L && \downarrow \L \\
... & \larroww Q & J^p\FF^{-1} & \larroww Q & J^p\FF^{0} & \larroww Q 
& J^p\FF^{1} & \larroww Q & J^p\FF^{2} & \larroww Q & ...
\earr
\label{complex1}
\ee
Here $J^p\FF^{g}$ are $DGRO(N,\oj)$ Fock modules, the vertical maps 
denote the module action, $Q^2=0$, and all squares commute. In this 
situation, $DGRO(N,\oj)$ will act in a well-defined manner on the 
cohomology groups $H^g(Q)$, which thus acquire a module structure.

The problem is now to find such a complex. A natural candidate is
found in the physics of gauge theories, as formulated cohomologically
in the anti-field formalism \cite{HT92}. 
The goal of classical physics is to find the stationary surface $\Sigma$,
i.e. the set of solutions to the Euler-Lagrange (EL) equations, 
viewed as a submanifold embedded in configuration space $\QQ$. Dually,
one wants to construct the function algebra $C(\Sigma) = C(\QQ)/\II$,
where $\II$ is the ideal generated by the EL equations. For each field
$\fa$ and EL equation $\Ea=0$, introduce an anti-field $\fsa$ of 
opposite Grassmann parity. The extended configuration space 
$C(\QQ^*)$ can be decomposed into subspaces $C^g(\QQ^*)$ 
of fixed antifield number $g$, where $\afn\fa = 0$, $\afn\fsa = 1$.
As is well known, the Koszul-Tate (KT) complex 
\be
0 \larroww \dlt C^0(\QQ^*) \larroww \dlt C^1(\QQ^*) \larroww \dlt 
C^2(\QQ^*) \larroww \dlt \ldots,
\label{complex2}
\ee
where $\dlt\fa = 0$ and $\dlt\fsa = \Ea$, yields a resolution of 
$C(\Sigma)$; the cohomology groups $H^g(\dlt) = 0$ unless $g=0$, and 
$H^0(\dlt) =C(\QQ)/\II$ \cite{HT92}.

The idea in this paper is to consider not just functions on the 
stationary surface, but all differential operators on it. The KT 
differential $\dlt$ can then be written as a bracket: $\dlt F = [Q,F]$, 
where the KT charge $Q = \int \Ea\psa$ and $\psa$ is the canonical 
momentum corresponding to $\fsa$. 
If we pass to the space of $p$-jets before momenta are introduced, 
the Fock construction applies. Since the
KT charge consists of commuting operators, it does not need to be 
normal ordered, and we recover precisely the situtation in
(\ref{complex1}); the cohomology groups are well-defined $DGRO(N,\oj)$
modules of lowest-energy type.

An outstanding
problem is to take the jet order $p$ to infinity, because
infinite jets essentially contain the same information as the original
fields. This limit is problematic, because the abelian charges diverge
with $p$. However, it was noted in \cite{Lar01a} that if we have several
independent jets, of order $p$, $p-1$, ..., $p-r$, we can arrange so
that the leading terms cancel, and the abelian charges are finite
in $N=r$ dimensions (they vanish in $N<r$ dimensions). This 
situation applies here, because the anti-fields correspond to lower-order
jets; the order depends on the order of the EL equations. A set of
consistency conditions can therefore be formulated. These conditions are
very restrictive, and natural solutions exist in four dimensions only.

Hence quantum diffeomorphism symmetry is only possible provided that
spacetime is four-dimen\-sional.

\section{DGRO algebra}
Let $\xi=\xmu(x)\dmu$, $x\in\RR^N$, $\dmu = \d/\d x^\mu$,
be a vector field, with commutator 
$[\xi,\eta] \equiv \xmu\dmu\ynu\dnu - \ynu\dnu\xmu\dmu$.
Greek indices $\mu,\nu = 1,2,..,N$ label the 
spacetime coordinates and the summation convention is used on all kinds 
of indices.
The diffeomorphism algebra (algebra of vector fields, Witt algebra) 
$\vect(N)$ is generated by Lie derivatives $\Lxi$.
In particular, we refer to diffeomorphisms on the circle as 
repara\-metri\-zations. They form an additional $\vect(1)$ algebra with 
generators $L_f$, where $f = f(t)d/dt$, $t\in S^1$, is a vector field 
on the circle.
Let $\map(N,\oj)$ be the current algebra corresponding to the
finite-dimen\-sional semisimple Lie algebra $\oj$ with basis $J^a$, 
structure constants $f^{ab}{}_c$, and Killing metric $\dlt^{ab}$. 
The brackets in $\oj$ are $[J^a, J^b] = if^{ab}{}_c J^c$.
A basis for $\map(N,\oj)$ is given by $\oj$-valued functions $X=X_a(x)J^a$ 
with commutator $[X,Y]=if^{ab}{}_c X_aY_bJ^c$.
Finally, let $Obs(N)$ be the space of local functionals of the
{\em observer's tractory} $\qmu(t)$, i.e. polynomial functions of  
$\qmu(t)$, $\dot\qmu(t)$, ... $d^k \qmu(t)/dt^k$,  $k$ finite, 
regarded as a commutative algebra. $Obs(N)$ is a $\vect(N)$ module in a 
natural manner.

The DGRO algebra $DGRO(N,\oj)$ is an abelian but non-central Lie algebra 
extension of $\vect(N) \ltimes \map(N,\oj) \oplus \vect(1)$ by $Obs(N)$:
\[
0 \longrightarrow Obs(N) \longrightarrow DGRO(N,\oj) \longrightarrow
 \vect(N)\ltimes \map(N,\oj)\oplus \vect(1) \longrightarrow 0.
\]
The brackets are given by
\bes
[\Lxi,\Leta] &=& \L_{[\xi,\eta]} 
 + {1\/2\pi i}\int dt\ \dot\qrho(t) 
 \Big\{ c_1 \drho\dnu\xmu(q(t))\dmu\ynu(q(t)) +\nl
 \bl+ c_2 \drho\dmu\xmu(q(t))\dnu\ynu(q(t)) \Big\}, \nl
{[}\Lxi, \J_X] &=& \J_{\xi X}, \nl
{[}\J_X, \J_Y] &=& \J_{[X,Y]} - {c_5\/2\pi i}\dlt^{ab}
 \int dt\ \dot\qrho(t)\drho X_a(q(t))Y_b(q(t)), \nl
{[}L_f, \Lxi] &=& {c_3\/4\pi i} \int dt\ 
 (\ddot f(t) - i\dot f(t))\dmu\xmu(q(t)), 
\label{DGRO} \\
{[}L_f,\J_X] &=& 0, \nl
{[}L_f,L_g] &=& L_{[f,g]} 
 + {c_4\/24\pi i}\int dt (\ddot f(t) \dot g(t) - \dot f(t) g(t)), \nl
{[}\Lxi,\qmu(t)] &=& \xmu(q(t)), \nl
{[}L_f,\qmu(t)] &=& -f(t)\dot\qmu(t), \nl
{[}\J_X, \qmu(t)] &=& {[}\qmu(s), \qnu(t)] = 0,
\eens
extended to all of $Obs(N)$ by Leibniz' rule and linearity.
The numbers $c_1-c_5$ are called {\em abelian charges}. In previous papers,
I considered a slightly more complicated extension which depends on three
additional abelian charges $c_6-c_8$, but they vanish automatically when
$\oj$ is semisimple. The DGRO algebra is the natural higher-dimensional
generalization of the Virasoro and affine Kac-Moody algebras.

\section{Koszul-Tate cohomology}

\subsection{Classical representations of $\dmap$}
Let $J^a = (J^{\al a}_\bt)$ and $T^\mu_\nu = (T^{\al\mu}_{\bt\nu})$ be
matrices satisfying $\oj$ and $gl(N)$, respectively, where the
brackets in $gl(N)$ are
\be
[T^\mu_\nu, T^\rho_\si] = 
 \dlt^\rho_\nu T^\mu_\si - \dlt^\mu_\si T^\rho_\nu.
\label{glN}
\ee
It is straightforward to verify that $\Lxi = \xmu\dmu + \dnu\xmu T^\nu_\mu$
and $\J_X = X_a(x)J^a$ satisfy $\dmap$. This implies that its modules are
tensor densities valued in $\oj$ modules. The $\dmap$ action is given by
\bes
[\Lxi, \fa(x)] &=& -\xmu(x)\dmu\fa(x)
- \dnu\xmu(x)T^{\bt\nu}_{\al\mu}\fb(x), \nle
{[}\J_X, \fa(x)] &=& -X_a(x)J^{\bt a}_\al\fb(x), 
\eens
Let $\QQ=\QQ(\phi)$ denote the module spanned by all $\fa(x)$, $x\in\RR^N$. 
In physics terms, $\QQ$ is our configuration space.

\subsection{ KT complex for functions of $x$}
Clearly, $\dmap$ acts not only on $\QQ$ but also on the space of local
functionals on $\QQ$; denote this space $C(\QQ)$. This module
is highly reducible; eg., $C(\QQ) = \oplus_{n=0}^\infty C_n(\QQ)$, where
$C_n(\QQ)$ consists of functionals that are homogeneous of degree $n$ in 
$\phi$.

An interesting submodule of $C(\QQ)$ can be constructed as follows.
Let $S=\int \dNx\ \pounds(\phi)$ be an invariant action (in the sense of 
physics) and $\pounds(\phi)$ the associated Lagrangian. The Lagrangian is 
a local functional of $\phi$, i.e. a function of $\fa(x)$ and its 
derivatives $\dmu\fa(x)$, $\dmu\dnu\fa(x)$, etc., up to some finite
order, all evaluated at the same point $x$. In practice, the Lagrangian
only depends on first-order derivatives. The Euler-Lagrange (EL) equations,
\be
\Ea(x) \equiv {\dlt S\/\dlt\fa(x)} 
= {\d\pounds\/\d\fa}(x) - {\dmu}{\d\pounds\/\dmu\fa}(x) = 0,
\label{EL}
\ee
generate an ideal $\II\subset C(\QQ)$, and the
factor space $C(\QQ)/\II$ is still a $\dmap$ module due to the invariance
assuption. This factor space is
most conveniently described as a resolution of a certain Koszul-Tate (KT)
complex. For each field $\fa(x)$, introduce an antifield $\fsa(x)$ 
transforming as the corresponding EL equation $\Ea(x)$. We then consider
the extended configuration space $\QQ^*$ as the span of $\fa(x)$ and 
$\fsa(x)$. Now consider the space of local functionals on $\QQ^*$:
$C(\QQ^*) = C(\QQ)\otimes C(\fs)$,
where $C(\fs)$ denotes the space of local functionals of $\fs(x)$.
If $\phi$ is bosonic ($C(\QQ)$ consists of symmetric functionals), then
$\fs$ is fermionic ($C(\fs)$ consists of anti-symmetric functionals), and 
vice versa. 

Define the {\em anti-field number} by $\afn\fa = 0$, $\afn\fsa = 1$.
$C(\QQ^*)$ can be decomposed into subspaces $C^g(\QQ^*)$ of fixed antifield
number $g$:
\be
C(\QQ^*) = \oplus_{g=0}^\infty C(\QQ)\otimes C^g(\fs)
\equiv \oplus_{g=0}^\infty C^g(\QQ^*).
\ee
The KT complex takes the form
\be
0 \larroww \dlt C^0(\QQ^*) \larroww \dlt C^1(\QQ^*) \larroww \dlt 
C^2(\QQ^*)\larroww \dlt \ldots
\ee
where the KT differential $\dlt$ is defined by
\be
\dlt\fa(x) = 0, \qquad
\dlt\fsa(x) = \Ea(x).
\ee
By a standard argument \cite{HT92}, the cohomology groups $H^g(\dlt) = 0$
unless $g=0$, and $H^0(\dlt) =C(\QQ)/\II$. $H^0(\dlt)$ can be thought of as
the space $C(\Sigma)$ of functions on the {\em stationary surface}
$\Sigma$, i.e. the set of solutions to the EL equations, embedded as
a submanifold in $\QQ$.

Introduce canonical momenta $\pa(x)=\dlt/\dlt\fa(x)$ and 
$\psa(x)=\dlt/\dlt\fsa(x)$ satisfying
\be
[\pa(x),\fb(y)] = [\psb(x), \fsa(y)] = \dlt^\al_\bt\dlt(x-y), 
\ee
and all other brackets vanish.
The antifield number is $\afn\pa = 0$, $\afn\psa = -1$.
The KT differential can then be written as a bracket: 
$\dlt F = [Q, F]$, where
\be
Q = \int \dNx\ \Ea(x)\psa(x).
\label{Q0}
\ee
Let $\PP$ be the phase space corresponding to $\QQ$, i.e. the span of
$\fa(x)$ and $\pa(x)$, and let $\PP^*$ be the enlarged phase space, i.e.
the span of $\fa(x)$, $\fsa(x)$, $\pa(x)$ and $\psa(x)$.
The expression (\ref{Q0}) defines a differential, also denoted by $Q$, 
which acts on the space $C(\PP^*)$ of local functionals on $\PP^*$. 
Note that
$C(\PP^*)$ is a non-commutative algebra, which can be thought of as the
algebra of differential operators on $\QQ^*$. The decomposition into
subspaces of fixed antifield number now extends indefinitely in both 
directions:
\be
C(\PP^*) = \oplus_{g=-\infty}^\infty C^g(\PP^*).
\ee
Accordingly, we obtain the two-sided complex
\be
\ldots \larroww Q C^{-1}(\PP^*) \larroww Q C^0(\PP^*)
\larroww Q C^1(\PP^*) \larroww Q \ldots
\label{complex3}
\ee
The cohomology group $H^0(Q)$ can be thought of as the space of 
differential operators on the stationary surface $\Sigma$. However, I
do not know if (\ref{complex3}) is a resolution, i.e. if the other
cohomology groups vanish.

There is a problem: the EL equations may be dependent, i.e. there 
may be relations of the form
\be
r^a(x) = r^a_\al(x)\Ea(x) \equiv 0,
\label{ra}
\ee
where $r^a_\al(x)$ is some functional of $\fa(x)$.
Then $H^1(Q) \neq 0$, because $r^a_\al(x)\fsa(x)$ is KT closed:
$[Q, r^a_\al(x)\fsa(x)] = 0$. The standard way to kill this unwanted 
cohomology is to introduce a second-order antifield $b^a(x)$. Let
$[Q, b^a(x)] = r^a_\al(x)\fsa(x)$, which makes the latter expression
exact and thus makes it vanish in cohomology. To obtain the explicit 
expression for $Q$, introduce the second-order antifield momentum
$c_a(x)$, with the non-zero bracket
$[c_a(x), b^b(x)] = \dlt^b_a \dlt(x-y)$. The full KT differential is
now
\be
Q = \int \dNx\ (\Ea(x)\psa(x) + r^a_\al(x)\fsa(x)c_a(x) ).
\label{Q}
\ee
There can in principle be relations also among the $r^a_\al(x)$ of the
form $Z^A(x) = Z^A_a(x) r^a_\al(x) \equiv 0$. If so, it is necessary to
introduce higher-order antifields to eliminate the unwanted cohomology.
However, we will assume that the gauge symmetries are irreducible, i.e.
that no non-trivial higher-order relations exist, since this is the case
in all experimentally established theories of physics.

The situation is summarized in the following table:
\be
\barr{|c|c|c|l|}
\hline
g &  \hbox{Field} & \hbox{Momentum} & \hbox{Ideal} \\
\hline
0 & \fa(x) &\pa(x) & - \\
1 & \fsa(x) &\psa(x) & \Ea(x)\approx 0 \\
2 & b^a(x) &c_a(x) & r^a_\al(x)\fsa(x) \approx 0 \\
\hline
\earr
\ee

\subsection{ KT complex for functions of $x$ and $t$}
In \cite{Lar98} Fock representations of $DGRO(N,\oj)$ were constructed. 
Its classical modules consist of $\oj$-valued tensor fields which also 
transform as densities under reparametrizations:
\bes
[\Lxi, \fa(x,t)] &=& -\xmu(x)\dmu\fa(x,t)
- \dnu\xmu(x)T^{\bt\nu}_{\al\mu}\fb(x,t), \nl
{[}\J_X, \fa(x,t)] &=& -X_a(x)J^{\bt a}_\al\fb(x,t), \\
{[}L_f, \fa(x,t)] &=& -f(t)\d_t\fa(x,t) 
- \la(\dot f(t)-if(t))\fa(x,t).
\eens
Denote the linear span of $\fa(x,t)$ by $\QQ(t)$ and the corresponding 
phase space by $\PP(t)$.
The EL equations now read $\Ea(x,t)=0$; they are obtained from (\ref{EL})
by replacing $\fa(x)$ by $\fa(x,t)$ everywhere.
The KT charge (\ref{Q}) is replaced by
\be
Q = \int \dxdt\ (\Ea(x,t)\psa(x,t) + r^a_\al(x,t)\fsa(x,t)c_a(x,t) ).
\ee
Since the space $C(\QQ(t))$ is larger than $C(\QQ)$, we must 
factor out a larger ideal to obtain a resolution of the same space
$C(\Sigma)$. It is easy to see that the necessary additional requirement 
is $\d_t\fa(x,t)\approx0$; to implement this 
constraint in cohomology, we introduce the antifield $\wfa(x,t)$ with
canonical momentum $\wpa(x,t)$. Since $\Ea(x,t)$ depends on $\fa(x,t)$
only, we now have $\d_t\Ea(x,t)=0$, which generates unwanted cohomology.
This is eliminated by introducing a second-order antifield $\wfsa(x,t)$.
Finally, the other second-order antifield $b^a(x,t)$, associated with
the gauge symmetry, is now reducible. Correct this by introducing a 
third-order antifield $\wba(x,t)$. 
The situation is summarized in the following table:
\be
\barr{|c|c|c|l|}
\hline
g & \hbox{Field} & \hbox{Momentum} & \hbox{Ideal} \\
\hline
0 & \fa(x,t)& \pa(x,t) & - \\
1 & \fsa(x,t) & \psa(x,t) & \Ea(x,t)\approx 0 \\
1 & \wfa(x,t)& \wpa(x,t) & \d_t\fa(x,t) \approx 0 \\
2 & b^a(x,t)& c_a(x,t) & r^a_\al(x,t)\fsa(x,t) \approx 0 \\
2 & \wfsa(x,t)& \wpsa(x,t) & \d_t\fsa(x,t) \approx 0 \\
3 & \wba(x,t)& \wca(x,t) & \d_t b^a(x,t) \approx 0 \\
\hline
\earr
\ee
The full KT differential becomes
\bes
Q &=& \int \dxdt\ (\Ea(x,t)\psa(x,t) + r^a_\al(x,t)\fsa(x,t)c_a(x,t) \\
&&+ \d_t\fa(x,t)\wpa(x,t) + \d_t\fsa(x,t)\wpsa(x,t) 
+ \d_t b^a(x,t)\wca(x,t)).
\eens
As before, we obtain resolutions of $C(\QQ(t))/\II(t) = C(\Sigma)$ and
$C(\PP(t))/\II(t)$, where $\II(t)$ is the totality of all relevant ideals.

\subsection{KT complex in jet space}
\label{ssec:jet}
The crucial idea in \cite{Lar98} is to introduce a priviledged curve
$\qmu(t)\in\RR^N$ in spacetime (``the observer's trajectory'') and to
expand all fields and antifields in a Taylor series around this curve,
before introducing canonical momenta. Hence e.g.,
\be
\fa(x,t) = \summ p {1\/\mm!} \fam(t)(x-q(t))^\mm,
\label{Taylor}
\ee
where $\mm = (m_1, \ab m_2, \ab ..., \ab m_N)$, all $m_\mu\geq0$, is a 
multi-index of length $|\mm| = \sum_{\mu=1}^N m_\mu$ and
$\mm! = m_1!m_2!...m_N!$.
Denote by $\mu$ a unit vector in the $\mu\th$ direction, so that
$\mm+\mu = (m_1, \ab ...,m_\mu+1, \ab ..., \ab m_N)$, and let
\be
\fam(t) = \d_\mm\fa(q(t),t)
= \underbrace{\d_1 .. \d_1}_{m_1} .. 
\underbrace{\d_N .. \d_N}_{m_N} \fa(q(t),t)
\label{jetdef}
\ee
be the $|\mm|\th$ order derivative of $\fa(x,t)$ on the
observer's trajectory $\qmu(t)$. Such objects transform as
\bes
[\Lxi, \fam(t)] &=& \d_\mm([\Lxi,\fa(q(t),t)]) 
+ [\Lxi,\qmu(t)]\dmu\d_\mm\fa(q(t),t) \nl
&\equiv& -\sumnm T^{\bt\nn}_{\al\mm}(\xi(q(t))) \fbn(t), \nl
{[}\J_X, \fam(t)] &=& \d_\mm([\J_X,\fa(q(t),t)]) 
\label{jet} \\
&\equiv& -\sumnm J^{\bt\nn}_{\al\mm}(X(q(t))) \fbn(t), \nl
{[}L_f, \fam(t)] &=& -f(t)\dot\fam(t) 
- \la(\dot f(t)-if(t))\fam(t),
\eens
where
\bes
&&T^\mm_\nn(\xi) \equiv (T^{\al\mm}_{\bt\nn}(\xi)) \nl
&&={\nn\choose\mm} \d_{\nn-\mm+\nu}\xmu T^\nu_\mu 
 + {\nn\choose\mm-\mu}\d_{\nn-\mm+\mu}\xmu
  - \dlt^{\mm-\mu}_\nn \xmu,
\label{Tmn}\\
&&J^\mm_\nn(X) \equiv (J^{\al\mm}_{\bt\nn}(X))
= {\nn\choose\mm} \d_{\nn-\mm} X_a J^a,
\eens
and
\be
{\mm\choose\nn} = {\mm!\/\nn!(\mm-\nn)!} = 
{m_1\choose n_1}{m_2\choose n_2}...{m_N\choose n_N}.
\ee
Here and henceforth we use the convention that a sum over
a multi-index runs over all values of length at most $p$. Since
$T^\nn_\mm(\xi)$ and $J^\nn_\mm(X)$ vanish whenever $|\nn|>|\mm|$,
the sums over $\nn$ in (\ref{jet}) are in fact further restricted.

Denote the space spanned by $\qmu(t)$ and $\{\fam(t)\}_{|\mm|\leq p}$
by $J^p\QQ$. $\fam(t)$ will be referred to as a $p$-jet, where $p$ is the
trunctation order\footnote{$p$-jets are usually defined as an equivalence
class of functions: two functions are equivalent if all derivatives up to
order $p$, evaluated at $\qmu$, agree. However, each class has a unique
representative which is a polynomial of order at most $p$, namely the
Taylor expansion around $\qmu$, so we may canonically identify jets with
Taylor series. Since $\qmu(t)$ depends on a parameter $t$, we deal in fact
with trajectories in jet space, but these will also be called jets for
brevity.}. This space is not a $DGRO(N,\oj)$ module, because 
diffeomorphisms act non-linearly on the trajectory, as can be seen in
(\ref{DGRO}). However, the space $C(J^p\QQ)$ of functionals on $J^p\QQ$
(local in $t$) {\em is} a module, because the action on a $p$-jet can never
produce a jet of order higher than $p$. Equivalently, there is a 
non-linear realization of the DGRO algebra on the jet space $J^p\QQ$.

Expand also the EL equations and the anti-fields in multi-dimensional 
Taylor series. Set 
$\Eam(t) = \d_\mm\Ea(q(t),t)$ and $\fsam(t) = \d_\mm\fsa(q(t),t)$.
What must be noted is that we can only define $\Eam(t)$ for 
$|\mm|\leq p-\ord_\al$, where $\ord_\al$ is the order of the EL equation 
$\Ea(x)$. This is because $\Eam(t)$ is a function of $\fan(t)$ for all
$|\nn| \leq |\mm|+\ord_\al$, and $\fan(t)$ is undefined for $|\nn|>p$.
Similarly, the relations (\ref{ra}) and the corresponding
second-order anti-fields $b^a(x)$ give rise to the jets
$\ram(t) = \d_\mm(r^a_\al(q(t),t) \fsa(q(t),t))$ and 
$\bam(t) = \d_\mm b^a(q(t),t)$, respectively. 
If the relations $r^a_\al$ are of order $\ordg_a$ in the derivatives, 
$\ram(t)$ and $\bam(t)$ is only defined for $|\mm|\leq p-\ordg_a$.

The conditions of type $\d_t\fa(x,t)$ give rise to additional constraints:
\bes
D_t\fam(t) &\equiv& \dot\fam(t) - \dot\qmu(t)\phi_{\al,\mm+\mu}(t)
\approx 0, \nl
D_t\fsam(t) &\equiv& \dot\fsam(t) - \dot\qmu(t)\fsa_{,\mm+\mu}(t)
\approx 0, 
\label{D_t}\\
D_t\bam(t) &\equiv& \dot\bam(t) - \dot\qmu(t)b^a_{,\mm+\mu}(t)
\approx 0.
\eens
These conditions are eliminated in cohomology by the introduction of
further (second and third order) anti-fields $\wfam(t)$, $\wfsam(t)$
and $\wbam(t)$. The conditions in (\ref{D_t}), and hence the barred
antifields, are only defined for one order less than the corresponding
unbarred antifield, since $|\mm+\mu| = |\mm|+1$.

Add dual coordinates (jet momenta) $\pmu(t)$, $\pam(t)$, $\psam(t)$, 
$\cam(t)$, $\wpam(t)$, $\wpsam(t)$ and $\wcam(t))$, which
satisfy
\bes
[\pmu(s), \qnu(t)] &=& \dlt^\nu_\mu \dlt(s-t), \nl
{[}\pam(s), \fbn(t)] &=& \dlt^\al_\bt\dlt^\mm_\nn \dlt(s-t), \nl
{[}\psam(s), \fsbn(t)] &=& \dlt_\al^\bt\dlt^\mm_\nn \dlt(s-t), \nl
{[}\cam(s), \bbn(t)] &=& \dlt^b_a \dlt^\mm_\nn \dlt(s-t), \\
{[}\wpam(s), \wfbn(t)] &=& \dlt^\al_\bt\dlt^\mm_\nn \dlt(s-t), \nl
{[}\wpsam(s), \wfsbn(t)] &=& \dlt_\al^\bt\dlt^\mm_\nn \dlt(s-t), \nl
{[}\wcam(s), \wbbn(t)] &=& \dlt^b_a \dlt^\mm_\nn \dlt(s-t),
\eens
and all other brackets vanish. 
Denote the phase space spanned by all jets and jet momenta by $J^p\PP^*$ and
the space of local functionals on $J^p\PP^*$ by $C(J^p\PP^*)$; alternatively,
this space may be considered as the differential operators on $J^p\QQ^*$.
The KT differential acting on the space of $C(J^p\PP^*)$ becomes
\bes
Q &=& \int dt\ \Big(\summ{p-\ord_\al}\Eam(t)\psam(t) 
+ \summ{p-\ordg_a}\ram(t)\cam(t) \nl
&&+ \summ{p-1}D_t\fam(t)\wpam(t) 
+ \summ{p-\ord_\al-1}D_t\fsam(t)\wpsam(t) \nl
&&+ \summ{p-\ordg_a-1}D_t\bam(x,t)\wcam(x,t) \Big).
\label{Qjet}
\ees
The situation is summarized in the following table:
\be
\barr{|c|c|c|c|c|}
\hline
g & \hbox{Field} & \hbox{Momentum} & \hbox{Order} &\hbox{Ideal} \\
\hline
0 & \fam(t)& \pam(t) & p & - \\
1 & \fsam(t) & \psam(t) & p-\ord_\al & \Eam(t)\approx 0 \\
1 & \wfam(t)& \wpam(t) & p-1 & D_t\fam(t) \approx 0 \\
2 & \bam(t)& \cam(t) & p-\ordg_a &\ram(t) \approx 0 \\
2 & \wfsam(t)& \wpsam(t) & p-\ord_\al-1 &D_t\fsam(t) \approx 0 \\
3 & \wbam(t)& \wcam(t) & p-\ordg_a-1 &D_t\bam(t) \approx 0 \\
\hline
\earr
\ee

\subsection{KT complex for the observer's trajectory}
\label{ssec:geo}
Until now we have not been very explicit about the set of fields.
In this subsection we will assume that among the fields $\fa(x)$ is
a metric $g_{\mu\nu}(x)$, either as a fundamental field or expressed 
in terms of vielbeine. We can then construct the following derived
quantites:
\begin{enumerate}
\item
The Levi-Civit\`a connection
$\Gamma^\nu_{\si\tau}(x,t) = \half g^{\nu\rho}(x,t)(\dsi g_{\rho\tau}(x,t)
+\break \dtau g_{\si\rho}(x,t) - \drho g_{\si\tau}(x,t))$.
\item
The einbein $e(t) = \sqrt{g_{\mu\nu}(q(t),t) \dot\qmu(t)\dot\qnu(t)}$.
\item
The repara\-metrization connection $\Gamma(t) = -e^{-1}(t)\dot e(t)$.
\end{enumerate}
Consider the geodesic operator
\be
\GG_\mu(t) =
e^{-1}(t) g_{\mu\nu}(t) ( \ddot\qnu(t) + \Gamma(t) \dot\qnu(t) 
+ \Gamma^\nu_{\si\tau}(t)\dot\qsi(t)\dot\qtau(t) ),
\label{geo}
\ee
where $g_{\mu\nu}(t)$ and $\Gamma^\nu_{\si\tau}(t)$ are the zero-jets
corresponding to the metric and Levi-Civit\`a connection, respectively.
If we define the proper time derivative by
\be
{d\phi\/d\tau}(t) = e^{-1}(t) {d\/dt}(e(t)\phi(t)),
\ee
(\ref{geo}) takes on the suggestive form
\be
\GG_\mu(t) = g_{\mu\nu}(t) ({d^2\qnu\/d\tau^2}(t)
 + \Gamma^\nu_{\si\tau}(t){d\qsi\/d\tau}(t){d\qtau\/d\tau}(t)).
\ee
It is straightforward to check that the geodesic equation 
$\GG_\mu(t) = 0$ transforms homogeneously under $DGRO(N,\oj)$:
\bes
{[}\Lxi, \GG_\nu(t)] &=& -\dnu\xmu(q(t))\GG_\mu(t), \nl
{[}\J_X, \GG_\nu(t)] &=& 0, \\
{[}L_f, \GG_\nu(t)] &=& -f(t)\dot\GG_\nu(t) - \dot f(t)\GG_\nu(t).
\eens
It can therefore be used to eliminate the observer's trajectory, apart 
from initial conditions. To implement this constraint in cohomology,
we introduce the trajectory antifield $q^*_\mu(t)$, with momentum
$p^{*\mu}(t)$. They obey the non-zero anticommutation relation
\be
[p^{*\mu}(s), q^*_\nu(t)] = \dlt^\mu_\nu \dlt(s-t),
\ee
which is fermionic since $\GG_\nu(t)$ is bosonic.
The contribution to the KT differential is
\be
Q = \int dt\ \GG_\mu(t) p^{*\mu}(t).
\label{QG}
\ee

\section{Quantization}
The KT complexes constructed in the previous section were all classical
in the sense that the abelian charges of the DGRO algebra vanish.
To quantize the theory, we introduce a Fock vacuum annihilated
by all negative Fourier modes; see \cite{Lar98} for an explicit 
description on how this is carried out. To avoid ill defined expressions
acting on the Fock vacuum, all expressions must be normal ordered with
respect to frequency; this is denoted by double dots ($\no{\quad}$).
It follows immediately from (\ref{jet}) that the following
operators define a realization of $DGRO(N,\oj)$ in Fock space:
\bes
\Lxi &=& \int dt\  \Big\{ \no{\xmu(q(t)) \pmu(t)} 
 + \sumnm T^{\bt\nn}_{\al\mm}(\xi(q(t))) \no{\fbn(t)\pam(t)} \Big\} \nl
&=& \int dt\ \Big\{ \no{\xmu(q(t)) \pmu(t)} 
 - \xmu(q(t)) P_\mu(t) +\nl
 &&+\summn p {\mm\choose \nn}
  \d_{\mm-\nn}\xmu(q(t)) E^\mm_{\nn+\mu}(t) \Big\}
 + T_{d\xi}, \nl
T_{d\xi} &=& \int dt\ \summn p {\mm\choose \nn} 
 \d_{\mm-\nn+\nu}\xmu(q(t)) T^{\mm\nu}_{\nn\mu}(t), 
\label{DROreal} \\
J_X  &=& \int dt\ \sumnm J^{\bt\nn}_{\al\mm}(X(q(t))) 
 \no{\fbn(t)\pam(t)} \nl
&=& \int dt\ \summn p {\mm\choose \nn} 
 \d_{\mm-\nn}X_a(q(t)) J^{\mm a}_\nn(t), \nl
L_f &=& \int dt\ f(t)L(t)+ \la(\dot f(t)-if(t))E(t),
\eens
where 
\bes
P_\mu(t) &=& \summ p E^\mm_{\mm+\mu}(t), \nl
L(t) &=& - \no{\dot\qmu(t)\pmu(t)} + F(t), \nl
E^\mm_\nn(t) &=& \no{ \pam(t)\fan(t) }, 
\nlb{Freal}
J^{\mm a}_\nn(t) &=& J^{\bt a}_\al \no{ \pam(t) \fbn(t) }, \nl 
T^{\mm\nu}_{\nn\mu}(t) &=& T^{\bt\mu}_{\al\nu} \no{ \pam(t) \fbn(t) }, \nl
F(t) &=& \summ p \no{\pam(t)\dot\fam(t)}
\eens
and $T^\mu_\nu = (T^{\al\mu}_{\bt\nu})$ and  $J^a = (J^{\al a}_\bt)$
are matrices that generate $gl(N)$ and $\oj$, respectively.
The currents in (\ref{Freal}) satisfy an algebra of the form
\bes
{[}T^{\mm\mu}_{\nn\nu}(s), T^{\rr\rho}_{\ss\si}(t)] &=& ...
+ {1\/2\pi i} (k_1\dlt^\mu_\si\dlt^\rho_\nu 
+ k_2\dlt^\mu_\nu\dlt^\rho_\si)
\dlt^\mm_\ss\dlt^\rr_\nn \dot\dlt(s-t), \nl
{[}T^{\mm\mu}_{\nn\nu}(s), E^\rr_\ss(t)] &=& ...
+ {k_3\/2\pi i} \dlt^\mu_\nu\dlt^\mm_\ss\dlt^\rr_\nn \dot\dlt(s-t), \nl
{[}T^{\mm\mu}_{\nn\nu}(s), J^{\rr b}_\ss(t)] &=& ... + 0, \nl
{[}J^{\mm a}_\nn(s), J^{\rr b}_\ss(t)] &=& ... 
+ {k_5\/2\pi i} \dlt^{ab}\dlt^\mm_\ss\dlt^\rr_\nn \dot\dlt(s-t), \nl
{[}J^{\mm a}_\nn(s), E^\rr_\ss(t)] &=& ... + 0, 
\nlb{KM}
{[}E^\mm_\nn(s), E^\rr_\ss(t)] &=&
... + {k_4\/2\pi i} \dlt^\mm_\ss\dlt^\rr_\nn \dot\dlt(s-t), \nl
{[}F(s),F(t)] &=& ...
 + {c\/24\pi i}\Np{}(\dddot\dlt(s-t) + \dot\dlt(s-t)), \nl
{[}F(s), E^\mm_\nn(t)] &=& ...
 + {d_0\/4\pi i} \dlt^\mm_\nn (\ddot\dlt(s-t)+i\dot\dlt(s-t)), \nl
{[}F(s), J^{\mm a}_\nn(t)] &=& ... + 0, \nl
{[}F(s), T^{\mm \mu}_{\nn\nu}(t)] &=& ...
+ {d_1\/4\pi i} \dlt^\mu_\nu \dlt^\mm_\nn(\ddot\dlt(s-t)+i\dot\dlt(s-t)).
\eens
Here I have not written down regular terms explicitly; they form an
open algebra which is described in \cite{Lar01a}.

Let $\rep$ be a $gl(N)$ representation and $M$ a $\oj$ representation.
Define numbers $u$,$v$,$w$,$x$,$y$ by
\bes
\barr{lll}
\tr\,1 = x, &\qquad&
\tr\,T^\mu_\nu T^\si_\tau = u\dlt^\mu_\tau\dlt^\si_\nu
+ v\dlt^\mu_\nu\dlt^\si_\tau, \\
\tr\,T^\mu_\nu = w\dlt^\mu_\nu, &\qquad&
\tr\,J^aJ^b = y\dlt^{ab},
\earr
\label{numdef}
\ees
where the trace is taken in the $\gg$ representation $M\oplus\rep$.
The relation to the numbers $k_0(\rep)$, $k_1(\rep)$, $k_2(\rep)$, 
and $y_M$ defined in \cite{Lar98} is
\bes
u = k_1(\rep)\, \dim\,M, &\qquad&
x = \dim\,\rep\,\dim\,M, \nl
v = k_2(\rep)\, \dim\,M, &\qquad&
y = \dim\,\rep\,y_M, \\
w = k_0(\rep)\, \dim\,M, &&
\eens
provided that $\rep$ and $M$ are irreducible.
The values of the abelian charges were given in \cite{Lar98}, Theorems
1 and 3, and again in \cite{Lar01a}, Theorem 1. They depend on the
central extensions in (\ref{KM}):
\bes
c_1 &=& 1 - k_1\Np{} - k_4 {N+p+1\choose N+2}, \nl
c_2 &=& - k_2\Np{} - 2k_3 \Np{+1} - k_4 {N+p\choose N+2}, \nl
c_3 &=& 1 + d_1\Np{} + d_0\Np{+1}, 
\label{cs}\\
c_4 &=& 2N - c\Np{}, \nl
c_5 &=& k_5\Np{}.
\eens
For the Fock module, we have
\bes
&&k_1 = \mp u, \qquad k_2 = \mp v, 
\qquad k_3 = \mp w, \qquad  k_4 = \mp x, \nle
&&k_5 = \mp y, \qquad d_0 = \mp x, \qquad d_1 = \mp w, \qquad  c = \mp x.
\eens
The upper signs apply to bosons and the lower signs to fermions.
Note that I have used the assumption that $\oj$ is
semisimple to put $k_6=k_7=k_8=0$. 

The $DGRO(N,\oj)$ representations obtained in this fashion are well 
defined for all finite values of the jet order $p$. In order 
to reconstruct the original field by means of the Taylor series
(\ref{Taylor}), one must take the limit $p\to\infty$. A necessary
condition for taking this limit is that the abelian charges have a
finite limit. 
Taken at face value, the prospects for succeeding
appear bleak. When $p$ is large, ${m+p\choose n} \approx p^n/n!$, so 
the abelian charges (\ref{cs}) diverge; the worst case is 
$c_1 \approx c_2 \approx p^{N+2}/(N+2)!$, which diverges in all dimensions
$N > -2$. In \cite{Lar01a} I devised a way out of this problem: consider 
a more general realization by taking the direct sum of operators 
corresponding to different values of the jet order $p$. Take
the sum of $r+1$ terms like those in (\ref{Freal}), with $p$ replaced by
$p$, $p-1$, ..., $p-r$, respectively, and with $\rep$ and $M$ replaced by
$\repi$ and $\Mi$ in the $p-i$ term. 

Such a sum of contributions arises naturally from the KT complex, because
the antifields are only defined up to an order smaller than $p$ (e.g.
$p-\ord_\al$ or $p-\ordg_a$). Summing the contributions from the various entries
in subsections \ref{ssec:jet} and \ref{ssec:geo}, (\ref{DROreal}) is
replaced by
\bes
\Lxi &=& \int dt\ \Big\{ \no{\xmu(q(t)) \pmu(t)} 
-\dnu\xmu(q(t))\no{q^*_\mu(t)p^{*\nu}(t)} + \nl
&&+\summn p T^{\bt\nn}_{\al\mm}(\xi(q(t))) \no{ \fbn(t)\pam(t)} +\nl
&&-\summn {p-\ord_\al} T^{\bt\nn}_{\al\mm}(\xi(q(t))) 
 \no{ \fsan(t)\psbm(t)}+\nl
&&+\summn {p-\ordg_a} T^{b\nn}_{a\mm}(\xi(q(t))) \no{ \ban(t)\cbm(t)} +\nle
&&+\summn {p-1} T^{\bt\nn}_{\al\mm}(\xi(q(t))) 
 \no{ \wfbn(t)\wpam(t)} +\nl
&&-\summn {p-\ord_\al-1} T^{\bt\nn}_{\al\mm}(\xi(q(t))) 
 \no{ \wfsan(t)\wpsbm(t)}+\nl
&&+\summn {p-\ordg_a-1} T^{b\nn}_{a\mm}(\xi(q(t))) 
 \no{ \wban(t)\wcbm(t)}
\Big\},
\eens
and similar contributions to $\J_X$ and $L_f$. Here 
$T^{\bt\nn}_{\al\mm}(\xi)$ and $T^{b\nn}_{a\mm}(\xi)$ are matrices in
the two different $gl(N)$ representions acting on fields and second-order
antifields, respectively; the action on first order antifields is dual to
the field action, because this is how the EL equations transform.

Denote the numbers $u,v,w,x,y$ in the modules $\repi$ and $\Mi$,
defined as in (\ref{numdef}), by $u_i,v_i,w_i,x_i,y_i$, 
respectively. Of course, there is 
only one contribution from the observer's trajectory. 
Then it was shown in \cite{Lar01a}, Theorem 3, that
\bes
c_1 = -U\Npr, &\qquad&
c_2 = -V\Npr, \nl
c_3 = W\Npr, &\qquad&
c_4 = X\Npr,
\label{finc} \\
c_5 = Y\Npr, 
\eens
where $u_0=U$, $v_0=V$, $w_0=W$, $x_0=X$ and $y_0=Y$, provided that the 
following conditions hold:
\bes
i&\qquad&u_i + \sum_{j=0}^{i-2}\sum_{\ell=0}^j x_\ell = \mri U, \nl
ii&\qquad&v_i + \sum_{j=0}^{i-1}(2w_j + \sum_{\ell=0}^{j-1} x_\ell) 
= \mri V, \nl
iii&\qquad&w_i + \sum_{j=0}^{i-1}x_j = \mri W, \nl
iv&\qquad&x_i = \mri X, 
\nlb{conds}
v&\qquad&y_i = \mri Y, \nl
vi&\qquad&\sum_{i=0}^r (2w_j + \sum_{\ell=0}^{i-1} x_\ell) = 0, \nl
vii&\qquad&\sum_{i=0}^r x_i = 0, \nl
viii&\qquad&\sum_{i=0}^{r-1}\sum_{j=0}^i x_j = 0.
\eens
The abelian charges diverge if $N>r$ and vanish if $N<r$. When $N=r$, 
the they are independent of $p$ and in general non-zero.

Define
\bes
\al_i &=& \sum_{j=0}^{i-2}\sum_{\ell=0}^j (-)^\ell {r\choose\ell}, \nl
\bt_i &=& \sum_{j=0}^{i-1} (-)^j {r\choose j},
\label{albt}\\
\gm_i &=& \sum_{j=0}^{i-1}\sum_{\ell=0}^{j-1}(-)^\ell {r\choose\ell}.
\eens
Using the recurrence formula
\be
{n\choose i} = {n\choose i-1} + {n-1\choose i-1}
\label{recur}
\ee
and a straightforward induction argument, it can be shown that
\be
\al_i = \gm_i = (-)^i {r-2\choose i-2}, \qquad
\bt_i = -(-)^i {r-1\choose i-1}.
\ee
The conditions (\ref{conds}) become
\bes
i&\qquad&u_i + \al_i X = \mri U, \nl
ii&\qquad&v_i + 2\bt_i W - \gm_i X = \mri V, \nl
iii&\qquad&w_i + \bt_i X = \mri W, 
\label{conds2}\\
iv&\qquad&x_i = \mri X, \nl
v&\qquad&y_i = \mri Y.
\eens
The remaining conditions follow immediately from the identities
$\al_{r+1} = \bt_{r+1} = 0$.

There is another, simpler way to arrive at (\ref{conds2}).
Ignoring the finite contributions from the observer's trajectory,
(\ref{cs}) can be rewritten as
\bes
c_1 &=& u A(p) + x C(p-1), \nl
c_2 &=& v A(p) + 2w B(p-1) + x C(p-2), \nl
c_3 &=& -w A(p) + x B(p-1), 
\label{csp}\\
c_4 &=& x A(p), \nl
c_5 &=& -y A(p),
\eens
where
\bes
A(p) &=& \summ p 1 = {N+p\choose N}, \nl
B(p) &=& \summ p (m_1+1) = {N+p+1\choose N+1}, \\
C(p) &=& \summ p (m_1+1)(m_2+1) = {N+p+2\choose N+2}.
\eens
As discussed in \cite{Lar01a}, the numbers arise from sums over 
multi-indices length $|\mm|\leq p$ of certain components. The restriction
to finite length can be viewed as a regularization, with the nice
property that diffeomorphism invariance is preserved. Another possible
regularization is to introduce the {\em fugacity}
$\z = (\z_1, ..., \z_N)$, and consider the expressions
\bes
A(\z) &=& \sum_{|\mm|} \z^{\mm} = \prod_{i=1}^N {1\/1-\z^i}, \nl
B(\z) &=& \sum_{|\mm|} (m_1+1)\z^{\mm} 
= (\z_1{\d\/\d\z_1}+1) A(\z)
= {1\/1-\z_1}A(\z) \\
C(\z) &=& \sum_{|\mm|} (m_1+1)(m_2+1)\z^{\mm} 
= (\z_1{\d\/\d\z_1}+1)(\z_2{\d\/\d\z_2}+1) A(\z) \nl
&=& {1\/1-\z_1}{1\/1-\z_2}A(\z).
\eens
If we now put all $\z_i = \z$, we obtain
\be
A(\z) = {1\/(1-\z)^N}, \quad
B(\z) = {1\/(1-\z)^{N+1}}, \quad
C(\z) = {1\/(1-\z)^{N+2}}.
\ee
The limit $p\to\infty$ is replaced by $\z\to1$.

However, it is not $B(p)$ and $C(p)$ that appear in (\ref{csp}), but 
rather $B(p-1)$, $C(p-1)$, and $C(p-2)$. If 
\be
A(p) \sim \sum_{k=0}^p a_k \z^k \to A(\z)
\ee
then
\be
A(p-1) \sim  \sum_{k=0}^p a_{k-1} \z^k 
= \sum_{\ell=0}^{p-1} a_\ell\z^{\ell+1} \to \z A(\z).
\ee
More generally, $A(p-k) \sim \z^k A(\z)$.

Equation (\ref{csp}) contains the contributions from a single, fermionic
jet of order $p$. With several jets of order $p$, ..., $p-r$, the term
$uA(p)$ becomes $u_0A(p) + u_1A(p-1) + ... + u_rA(p-r)$, so we must 
replace the parameter $u$ with the function 
$u(\z) = \sum_{i=0}^r u_i \z^i$, 
Taking this into account, (\ref{csp}) corresponds to
\bes
c_1 &=& u(\z) A(\z) + x(\z) \z C(\z), \nl
c_2 &=& v(\z) A(\z) + 2w(\z) \z B(\z) + x(\z) \z^2C(\z), \nl
c_3 &=& -w(\z) A(\z) + x(\z) \z B(\z), 
\label{csz}\\
c_4 &=& x(\z) A(\z), \nl
c_5 &=& -y(\z) A(\z).
\eens
Now demand that
\be
c_1 = U, \qquad c_2 = V, \qquad c_3 = -W, \qquad c_4 = X,
\qquad c_5 = -Y,
\label{cslim}
\ee
in the limit $\z\to1$. The simplest way to achieve this is to require
that (\ref{cslim}) holds for all $\z$. This leads to
\bes
u(\z) &=& U(1-\z)^N - X\z (1-\z)^{N-2}, \nl
v(\z) &=& V(1-\z)^N + 2W\z (1-\z)^{N-1} - X\z^2 (1-\z)^{N-2}, \nl
w(\z) &=& W(1-\z)^N - X\z (1-\z)^{N-1}, \\
x(\z) &=& X(1-\z)^N, \nl
y(\z) &=& Y(1-\z)^N.
\eens
Equation (\ref{conds2}) is recovered after an expansion in $\z$. E.g., 
\be
x(\z) = \sum_{i=0}^r x_i\z^i = X(1-\z)^N 
= \sum_{i=0}^N (-)^i {N\choose i}X\z^i,
\ee
leading to $x_i = (-)^i {N\choose i}X$ and $r=N$.

\section{Solutions to the constraint equations}
\subsection{Original constraint equations}
Let us now consider the solutions of (\ref{conds}) for the numbers
$x_i$, which can be interpreted as the number of fields and anti-fields.
First assume
that the field $\fam(t)$ is fermionic with $x_F$ components, which gives
$x_0=x_F$. We may assume, by the spin-statistics theorem, that the 
EL equations are first order, so the bosonic antifields $\fsam(t)$ 
contribute $-x_F$ to $x_1$. The barred antifields $\wfam(t)$ are also
defined up to order $p-1$, and so give $x_1=-x_F$, and the barred 
second-order antifields $\wfsam(t)$ give $x_2=x_F$. Further assume that
the fermionic EL equations have $x_S$ gauge symmetries, i.e. the 
second-order antifields $\bam(t)$ give $x_2=x_S$. In established
theories, $x_S=0$, but we will need a non-zero value for $x_S$.
Finally, the corresponding barred antifields give $x_3 = -x_S$.

For bosons the situation is analogous, with two exceptions: all signs
are reversed, and the EL equations are assumed to be second order.
Hence $\fsam(t)$ yields $x_2=x_B$ and the gauge antifields $\bam(t)$ 
give $x_3=-x_G$. Accordingly, the barred antifields are one order higher.

The situation is summarized in the following tables, where the upper half
is valid if the original field is fermionic and the lower half if it is
bosonic:
\bes
\barr{|c|c|c|l|}
\hline
g & \hbox{Field} & \hbox{Order} &x \\
\hline
0 & \fam(t) & p & x_F \\
1 & \wfam(t) & p-1 & -x_F \\
1 & \fsam(t) & p-1 & -x_F \\
2 & \wfsam(t) & p-2 & x_F \\
2 & \bam(t) & p-2 & x_S \\
3 & \wbam(t) & p-3 & -x_S \\
\hline
\hline
0 & \fam(t) & p & -x_B \\
1 & \wfam(t) & p-1 & x_B \\
1 & \fsam(t) & p-2 & x_B \\
2 & \wfsam(t) & p-3 & -x_B \\
2 & \bam(t) & p-3 & -x_G \\
3 & \wbam(t) & p-4 & x_G \\
\hline
\earr
\label{tab}
\ees
If we add all contributions of the same order, we see that relation 
$iv$ in (\ref{conds2}) can only be satisfied provided that
\bes
p: &\quad& x_F - x_B = X \nl
p-1: && -2x_F+x_B = -rX, \nl
p-2: && x_B + x_F + x_S = {r\choose2}X, \nl
p-3: && -x_B-x_S-x_G = -{r\choose3}X, 
\label{rcond}\\
p-4: && x_G = {r\choose4}X, \nl
p-5: && 0 = -{r\choose5}X, ...
\eens
The last equation holds only if $r\leq4$ (or trivially if $X=0$). 
On the other hand, if we demand that there is at least one bosonic
gauge condition, the $p-4$ equation yields $r\geq4$. Such a demand is
natural, because both the Maxwell/Yang-Mills and the Einstein equations
have this property. Therefore, we are unambigiously guided to consider
$r=4$ (and thus $N=4$). The specialization of (\ref{rcond}) to four
dimensions reads
\bes
p: &\quad& x_F - x_B = X \nl
p-1: && -2x_F+x_B = -4X, \nl
p-2: && x_B + x_F + x_S = 6X, \\
p-3: && -x_B-x_S-x_G = -4X, \nl
p-4: && x_G = X.
\eens
Clearly, the unique solution to these equations is
\be
x_F = 3X, \qquad x_B = 2X, \qquad x_S = X, \qquad x_G = X.
\label{xsol}
\ee

\subsection{Reduced constraint equations}
The barred antifields, associated with the constraints (\ref{D_t}), can
conveniently be eliminated first. 
For each field or antifield of order $p-\ord_\al$, there is a 
corresponding barred antifield of opposite Grassmann parity and one order 
lower. E.g., for the field $\fam(t)$ at order $p$, we have the antifield
$\wfam(t)$ at order $p-1$, and for $\fsam(t)$ at order $p-\ord_\al$, we
have $\wfsam(t)$ at order $p-\ord_\al-1$. Thus, if $x_i$ has a contribution
$x_i'$ from an unbarred (anti-)field, then $x_{i+1}$ has the contribution
$-x_i'$. In particular, $x_r' = 0$. 
This means that the total value for $c_4$ in (\ref{cs}) becomes
\bes
c_4 &=& \sum_{i=0}^r x_i {N+p-i\choose N} \nl
&=& \sum_{i=0}^{r-1} x_i' {N+p-i\choose N}
- \sum_{i=1}^r x_{i-1}' {N+p-i\choose N} \nle
&=& \sum_{i=0}^{r-1} x_i' 
 \Big\{ {N+p-i\choose N} - {N+p-i-1\choose N}\Big\} \nl
&=& \sum_{i=0}^{r-1} x_i'{N-1+p-i\choose N-1}.
\eens
We recognize that this expression is of the same form as the original
expression, with the replacements $x_i\to x_i'$, $N\to N-1$, and 
$r\to r-1$. The finiteness conditions for the original parameters $x_i$
are thus equivalent to the same conditions for the reduced parameters
$x_i'$ in one dimension less. Henceforth we only consider the reduced
parameters, and skip the primes to avoid unnecessarily cumbersome
notation.

The tables (\ref{tab}) are replaced by (fermions first, bosons second)
\bes
\barr{|c|c|c|l|}
\hline
g & \hbox{Field} & \hbox{Order} &x \\
\hline
0 & \fam(t) & p & x_F \\
1 & \fsam(t) & p-1 & -x_F \\
2 & \bam(t) & p-2 & x_S \\
\hline
\hline
0 & \fam(t) & p & -x_B \\
1 & \fsam(t) & p-2 & x_B \\
2 & \bam(t) & p-3 & -x_G \\
\hline
\earr
\ees
The reduced version of (\ref{rcond}) becomes
\bes
p: &\quad& x_F - x_B = X \nl
p-1: && -x_F = -rX, \nl
p-2: && x_B + x_S = {r\choose2}X, 
\label{rcond2}\\
p-3: && -x_G = -{r\choose3}X, \nl
p-4: && 0 = {r\choose4}X, ...
\eens
If we sum the first three equations, we obtain $x_S = (r^2 -3r + 2)X/2$.
Thus the only case where we can avoid the fermionic gauge symmetries
is if $r=2$, i.e. $N=3$. However, if we put $r=2$, we get from the
fourth equation $x_G = 0$, which means that there are no bosonic gauge
symmetries either. But this can not be the case, assuming that the
Einstein equation is included among our EL equations, and thus $r>2$.
However, the last equation is clearly impossible to satisfy if $r\geq4$.
For $r=3$, i.e. $N=4$, (\ref{rcond2}) becomes
\bes
p: &\quad& x_F - x_B = X \nl
p-1: && -x_F = -3X, 
\nle
p-2: && x_B + x_S = 3X, \nl
p-3: && -x_G = -X.
\eens
The solution to these equations is of course still given by (\ref{xsol}).

The solutions to the other conditions in (\ref{conds}) follow by analogous
considerations. Assume that there are $x_F$ fermions that contribute 
$x_F$, $y_F$, $u_F$, $v_F$, $w_F$ to parameters $x$, $y$, $u$, $v$, $w$,
respectively. The contributions from the $x_B$ bosons, $x_S$ fermionic
gauge conditions, and $x_G$ bosonic gauge conditions are denoted 
analogously. Since we have already excluded $N\neq4$, we limit
ourselves to this case. We also use the reduced parameters and hence
we set $r=3$. The numbers $\al_i$ and $\beta_i$ in (\ref{albt}) are
\bes
\barr{|c|c|c|c|}
\hline
i & \al_i & \beta_i & \gm_i \\
\hline
0 & 0 & 0 & 0\\
1 & 0 & 1 & 0\\
2 & 1 & -2 & 1\\
3 & -1 & 1 & -1\\
\hline
\earr
\ees
By following exactly the same arguments as for $x_i$ above, we see that
the reduced version of (\ref{conds2}) becomes
\bes
\barr{rllll}
p:  && x_F-x_B = X && y_F-y_B = Y \\
p-1:&& -x_F = -3X  && -y_F = -3Y \\
p-2:&& x_B+x_S=3X  && y_B+y_S=3Y \\
p-3:&& -x_G = -X   && -y_G=-Y \\
\\
p: && u_F-u_B=U     && v_F-v_B=V \\
p-1: && -u_F = -3U  && -v_F=-3V-2W \\
p-2: && u_B+u_S=3U-X&& v_B+v_S=3V+4W+X \\
p-3: && -u_G = -U+X && -v_G = -V-W-X \\
\\
p: && w_F-w_B=W\\
p-1: && -w_F=-3W-X\\
p-2: && w_B+w_S=3W+2X\\
p-3: && -w_G=-W-X \\
\earr
\ees
The solutions are given by
\bes
\barr{lllll}
x_B = 2X & y_B = 2Y \\
x_F = 3X & y_F = 3Y \\
x_S = X  & y_S = Y \\
x_G = X  & y_G = Y \\
\\
u_B = 2U  & v_B = 2V+2W \\
u_F = 3U  & v_F = 3V+2W \\
u_S = U-X & v_S = V+2W+X \\
u_G = U-X & v_G = V+2W+X \\
\\
w_B = 2W+X \\
w_F = 3W+X \\
w_S = W+X \\
w_G = W+X \\
\earr
\ees
This is our main result. It expresses the twenty parameters $x_B-w_G$
in terms of the five parameters $X$, $Y$, $U$, $V$, $W$.
For this particular choice of parameters, the abelian charges in
(\ref{finc}) are given by (\ref{cslim}),
independent of $p$. Hence there is no manifest obstruction to the
limit $p\to\infty$.

\subsection{Comparison with known physics}
\label{sec:QGSM4}
All experimentally known physics is well described by quantum theory, 
gravity, and the standard model in four dimensions.
We have already seen that quantum general covariance more or less 
dictates that spacetime has $N=4$ dimensions (\ref{rcond}). It is therefore
interesting to investigate to what extent the particle content 
matches (\ref{xsol}); recall that $x = \tr 1$ equals the number of field
components.

The bosonic content of the theory is given by the following table. Standard
notation for the fields is used, and one must remember that it is the 
na\"\i ve number
of components that enters the equation, not the gauge-invariant physical
content. E.g., the photon is described by the four components $A_\mu$ 
rather than the two physical transverse components.
\bes
\barr{|c|c|c|c|}
\hline
\hbox{Field} & \hbox{Name}& \hbox{EL equation} & x_B  \\
\hline
A^a_\mu & \hbox{Gauge bosons}
&D_\nu F^{a\mu\nu} = j^{a\mu} &12\times4 = 48  \\
g_{\mu\nu} & \hbox{Metric}
&G^{\mu\nu} = {1\/8\pi}T^{\mu\nu} & 10 \\
H & \hbox{Higgs field}
& g^{\mu\nu}\dmu\dnu H = V(H) & 2\\
\hline
\earr
\nle
\barr{|c|c|}
\hline
\hbox{Gauge condition} & x_G  \\
\hline
D_\mu D_\nu F^{a\mu\nu} = 0 & 12\times1 = 12 \\
\dnu G^{\mu\nu} = 0 & 4 \\
\hline
\earr
\eens
The total number of bosons in the theory is thus $x_B = 48+10+2 = 60$,
which implies $X=30$ by (\ref{xsol}). The number of gauge conditions is
$x_G = 16$, which implies $X=16$. There is certainly a discrepancy here.

The fermionic content in the first generation is given by
\bes
\barr{|c|c|c|c|}
\hline
\hbox{Field} & \hbox{Name}& \hbox{EL equation} & x_F  \\
\hline
u & \hbox{Up quark} & \Dslash u = ... & 2\times3 = 6\\
d & \hbox{Down quark} & \Dslash d = ... & 2\times3 = 6\\
e & \hbox{Electron} & \Dslash e = ... & 2\\
\nu_L & \hbox{Left-handed neutrino} & \Dslash \nu_L = ... & 1\\
\hline
\earr
\ees
The number of fermions in the first generation is thus 
$x_F = 6+6+2+1 = 15$. Counting all three generations and anti-particles,
we find that the total number of fermions is $x_F = 2\times3\times15 = 90$,
which implies $X=30$. There are no fermionic gauge conditions, so $x_S=0$,
which implies $X=0$.

It is clear that the predictions for $X$ ($30,16,30,0$) are not mutually
consistent. However, to cancel the leading terms, of order $p$ and $p-1$,
it is only necessary that $2x_F = 3x_B$, which is indeed the case in known
physics. It is therefore tempting to speculate that known physics is a
first approximation of a more elegant theory, which has the same field
content but more gauge conditions, including fermionic ones. This issue
will be addressed in the discussion.

It is important to check that the results remain the same if the same 
physical situation is described with a different, but equivalent, set of
fields. Typically, such spurious degrees of freedom have algebraic
EL equations. Denote the original (bosonic, say) $x_B$ fields by $\fam(t)$
and let $\psi_{i,\mm}(t)$ be $x_A$ spurious fields, defined for 
$|\mm|\leq p$. The contribution to $x_0$ from the bosonic fields is thus
$x_B + x_A$. There are also $x_A$ new EL equations $E^i_{,\mm}(t)$,
defined for $|\mm|\leq p$ because they are algebraic; $E^i_{,\mm}(t)$
contains $\psi_{j,\nn}(t)$ for all $|\nn|\leq|\mm|$, but not of higher
order. The corresponding anti-fields $\psi^{*i}_{,\mm}(t)$ add $-x_A$ to
$x_0$. The total result is $x_0 = x_B+x_A-x_A = x_B$, as before.

An example is given by the gravitational field in vielbein formalism.
Instead of the $10$ components of the metric $g_{\mu\nu} = g_{\nu\mu}$
we have the $16$ vielbeine $e^i_\mu$. However, the requirement that the 
metric $g_{\mu\nu} = e^i_\mu e_{i\nu}$ be symmetric gives rise to 
$6$ algebraic conditions, so the contribution to $x_0$ is still $10$.

\section{Discussion}

There are two key lessons to be learnt from twentieth century physics:
\begin{itemize}
\item
General relativity teaches us the importance of diffeomorphism invariance.
Physics is fully relational; there is no background stage {\em over}
which physics takes place. Rather, geometry itself participates actively
in the dynamics. Note that this is very different from mere coordinate
invariance, because there is no compensating background metric.
\item
Quantum theory teaches us the importance of projective lowest-energy
representations; the passage from Poisson brackets to commutators
makes normal ordering necessary.
E.g., to study angular momentum from a quantum perspective, it is not
sufficient to limit oneself to proper (integer spin) representations
of the rotation group $SO(3)$; one must also include the projective
(half-integer spin) representations. For finite-dimensional groups
such as $SO(3)$, projectivity only manifests itself on the group level,
but in the infinite-dimensional case already the Lie algebra is 
modified. Algebras of linear growth acquire central extensions, e.g.
the Virasoro and affine Kac-Moody algebras, whereas algebras of polynomial
but non-linear growth acquire abelian but non-central extensions.
\end{itemize}
The successful construction of a quantum theory of gravity will probably
combine these two insights. It seems obvious that the correct way to
combine diffeomorphism invariance and projective representations is to
consider projective representations of the diffeomorphism group, which
on the Lie algebra level gives rise to the DGRO algebra. To even think 
about quantum gravity without understanding $DGRO(N,\oj)$ seems to be 
a doomed project. It would be like doing 
classical gravity without tensor fields (= proper representations of 
the diffeomorphism group), or like doing quantum theory without
spinors (= projective representations of $SO(3)$).
Hence it seems appropriate to refer to DGRO algebra
symmetry as {\em quantum general covariance}.
In fact, the Fock modules considered in 
\cite{Lar98} automatically solve some of the outstanding problems in 
quantum gravity. By definition, it clarifies the
role of diffeomorphisms, and there are no causality problems, 
because the theory only involves events on the observer's trajectory and
such events are always causally related \cite{Lar01c}.

With the introduction of the Koszul-Tate cohomology in the present paper, 
dynamics has entered representation theory, presumably for the first time.
Since the classical KT cohomology is equivalent 
to standard formulations of classical physics, and the presence of 
Virasoro-like cocycles signals quantization, the 
construction in the present paper can be regarded as a novel 
quantization method, although the relation to other quantization schemes
is unclear. An important feature is that abelian extensions pose no
problem, as long as they are finite in the $p\to\infty$ limit. 

The existence of this limit may be viewed as a requirement on 
objective reality; the $p$-jets living on the observer's trajectory can 
be extended to fields defined throughout spacetime by means of a 
Taylor expansion, only if the limit $p\to\infty$ is well defined. 
This imposes severe constraints on the
field content. As we saw in Section \ref{sec:QGSM4}, most of these 
conditions are in qualitative agreement 
with established theories of physics, in particular with the standard 
model. We find both fermions and bosons, 
with EL equations of first and second order, respectively, and
bosonic gauge constraints of third order. 
Moreover, spacetime must have four dimensions provided that there are no
reducible gauge conditions. 

However, there is also sharp disagreement
on the number of gauge conditions. 
Quantum general covariance predicts the existence of fermionic gauge 
freedom of second order, and additional bosonic gauge freedom at third 
order. This points toward some kind of modification, maybe involving 
superalgebras.
An interesting possibility is to consider a gauge theory based
on the exceptional Lie superalgebra $\mb(3|8)$, which is the simple
vectorial superalgebra of maximal depth $3$ \cite{Kac98,Sh99}. 
Classically, the corresponding gauge
algebra $\map(N, \mb(3|8))$ acts on functions $\fa(x,y)$, valued in
modules of the grade zero subalgebra $sl(3)\oplus sl(2)\oplus gl(1)$,
i.e. the non-compact form of the symmetries of the standard model.
Here $x = (x^\mu) \in \RR^N$ is a spacetime coordinate and 
$y = (y^i) \in \CC^{3|8}$ is a coordinate in internal space. I have
recently attempted to generalize the standard model to a gauge theory
with $\mb(3|8)$ symmetry, and extra conditions on the fermionic fields
do indeed arise \cite{Lar02}. However, the considerations in that paper
are purely classical. It is clear that quantization in the spirit of the
present paper can be carried out 
($\map(N,\mb(3|8)) \subset \vect(N+3|8)$), but this task has not yet
been undertaken.

\end{document}